 \documentclass[final,5p,times,twocolumn]{elsarticle}

\usepackage{amssymb}

\usepackage{soul} \usepackage{multirow} \usepackage{bm}
\usepackage{array} \usepackage{amsmath} 
\usepackage{url} 

\journal{Computer Physics Communications}

\begin{document}

\begin{frontmatter}



\title{HSMA: An $O(N)$ electrostatics package implemented in LAMMPS}

\author[label1,label3]{Jiuyang Liang} \address[label1]{School of Mathematical
  Sciences, Shanghai Jiao Tong University, Shanghai 200240, China}
\ead{liangjiuyang@sjtu.edu.cn}

\author[label2]{Jiaxing Yuan} \address[label2]{School of Physics and
  Astronomy, Shanghai Jiao Tong University, Shanghai 200240, China}
\ead{yuanjiaxing123@hotmail.com}

\author[label1,label3]{Zhenli Xu\corref{author}} \address[label3]{Institute of
  Natural Sciences and MoE-LSC, Shanghai Jiao Tong University,
  Shanghai 200240, China}
\ead{xuzl@sjtu.edu.cn}

\cortext[author] {Corresponding author.}

\begin{abstract}
We implement two recently developed fast Coulomb solvers, 
HSMA3D [J. Chem. Phys. 149 (8) (2018) 084111] and HSMA2D [J. Chem. Phys. 152 (13) (2020) 134109], into a new 
user package HSMA
for molecular dynamics simulation engine LAMMPS.
The HSMA package is designed for efficient and accurate modeling
of electrostatic interactions in 3D and 2D periodic systems 
with dielectric effects at the $O(N)$ cost.
The implementation is hybrid MPI and OpenMP parallelized 
and compatible with existing LAMMPS functionalities.
The vectorization technique following AVX512 instructions is adopted for
acceleration. 
To establish the validity of our implementation, we have presented extensive 
comparisons to the widely used particle-particle particle-mesh (PPPM) algorithm
in LAMMPS and other dielectric solvers.
With the proper choice of algorithm parameters and parallelization setup, the package enables calculations of electrostatic interactions 
that outperform the standard PPPM 
in speed for a wide range of particle numbers.
\end{abstract}


\end{frontmatter}



{\bf PROGRAM SUMMARY/NEW VERSION PROGRAM SUMMARY}

\begin{small}
\noindent
{\em Program Title: HSMA} \\ {\em CPC Library link to program files:}
(to be added by Technical Editor) \\ {\em Developer's repository
  link:https://github.com/LiangJiuyang/HSMA-Harmonic-surface-mapping-algorithm-in-LAMMPS}
\\ {\em Code Ocean capsule:} (to be added by Technical Editor)\\ {\em
  Licensing provisions(please choose one):} GPLv3 \\ {\em Programming
  language: C++} \\
{\em Nature of problem: Evaluation of long-range electrostatic interactions for
  charged system with fully periodic condition or confined by planar
  dielectric interfaces.}\\
{\em Solution method: We implement the Harmonic Surface Mapping
  algorithm (HSMA), which combines the image-charge method with
  the harmonic surface mapping and converts the contribution of
  infinite images into a finite number of surface charges on an
  auxiliary sphere, into simulation package LAMMPS. 
  The HSMA package works for both fully periodic system and
  partially periodic system with planar dielectric interfaces, achieving
  truly linear $O(N)$ complexity by employing fast multiple method. Our package can be
  applied to general all-atom simulation and a broad range of charged
  complex fluids under dielectric confinement.  }\\
{\em Additional comments: Hybrid MPI plus OpenMP parallelization is
  used in HSMA package for high-performance
  computing. We provide vertor optimization by using the AVX512
  instructions and Intel MKL library for further acceleration. Intel
  parallel studio is required for these techniques. }\\
   \\

\end{small}

\section{Introduction}

Electrostatic interactions play a key role in nanoparticle self-assembly processes~\cite{wong00,Boles2016}, 
and are also so ubiquitous in biomolecular systems, such as DNA aggregation~\cite{burak03} and
polyelectrolyte complexation~\cite{Zhou2018,Lund2005a,lund13}.
Numerical simulations, such as Monte Carlo (MC) and molecular dynamics (MD), 
provide higher accuracy for understanding the long-range nature of the Coulomb interactions,
compared to theoretical efforts at the mean-ﬁeld level~\cite{borukhov97,silalahi10}.
In computer simulations, periodic boundary conditions (PBCs) are often employed 
for approximating an infinite system by using a small unit cell~\cite{frenkel-smit2}.
To efficiently sum over the contribution of charges in the infinite periodic replicates, 
the common approaches rely on the conventional Ewald summation [scaling with the particle number~$N$ as $O(N^{3/2})$]~\cite{ewald21} and its variations such as particle-particle particle-mesh Ewald (PPPM) [scaling $O(N\log(N))$]~\cite{hockney2021computer} and random batch Ewald (RBE) [scaling $O(N)$]~\cite{jin2020random}, or the multipole-type methods such as the tree code [scaling $O(N\log(N))$]~\cite{barnes1986hierarchical} and the fast multipole method (FMM) [scaling $O(N)$]~\cite{greengard1987fast}. 
Given their prevalence, the efficient parallel implementations~\cite{pollock1996comments,brown2012implementing,bush2006daft} in MD simulation packages such as LAMMPS~\cite{plimpton1995fast} and GROMACS~\cite{berendsen1995gromacs}
and the comparisons of these methods~\cite{arnold2013} have been the subject of considerable computational research efforts.

Furthermore, many materials interfaces are characterized by a discontinuity in the
permittivity, resulting in induced surface polarization charges.
This induced charge may significantly affect the ionic
distributions~\cite{messina04,barros14b,dossantos16b,wu18b,yuan20,ma2017ion} 
and dynamics~\cite{zhang11,antila18a,yuan19a},
so that developing efficient methods for modeling spatially varying dielectrics has received 
considerable research attention~\cite{boda04,barros14a,jiang16,wu18a,jadhao12,jadhao13,gan2016hybrid}. 
Whereas this is typically costly, the special situation of planar interfaces (as well as systems that can be approximated as such) can be treated highly efficiently and accurately via the well-known image-charge method (ICM)~\cite{thomson1848}. Still, in the presence of two planar interfaces, this leads to iterative image reflections, 
so that the summation over the infinite image charges is a non-trivial task
and several numerical methods such as ICM-MMM2D~\cite{tyagi07} (scaling $\mathcal{O}(N^{5/3})$), ectrostatic layer correction
method~\cite{tyagi08} (ICM-ELC, scaling as $\mathcal{O}(N\log N)$),
ICM-Ewald~\cite{dossantos15} (scaling as
$\mathcal{O}(N^{3/2})$), and ICM-PPPM (scaling as
$\mathcal{O}(N\log(N))$)~\cite{yuan2021} have been proposed. 

To benefit the community, we present a parallel implementation 
of two previously formulated HSMA methods~\cite{zhao2018,liang2020} into
the popular MD package LAMMPS~\cite{plimpton95},
producing highly efficient solvers for modeling electrostatic interactions in 3D and 2D periodic system. 
This article describes the new implementation and analyses its performance. 
The organization of this article is as follows. In Section~\ref{sec:simulation_model_HSMA}, we
present a description of the simulation model, with a brief review of our HSMA algorithms. 
Section~\ref{sec:lammps} provides the details of the LAMMPS implementation.
In Section~\ref{sec:accuracy}, we apply the method to three types of charged systems and confirm the correct implementation in LAMMPS: 
primitive electrolytes and SPC/E water in 3D bulk, and primitive electrolytes confined between parallel dielectric interfaces.
In Section~\ref{sec:CPU}, we analyze the CPU performance of the HSMA.

\section{Overview of simulation model and HSMA}
\label{sec:simulation_model_HSMA}

\subsection{Simulation model}

The package is designed for modeling two types of systems, 
namely 3D-periodic system
and 2D-periodic system with dielectric boundaries.
\begin{figure}[!ht]
	\centering
        \includegraphics[width=\linewidth]{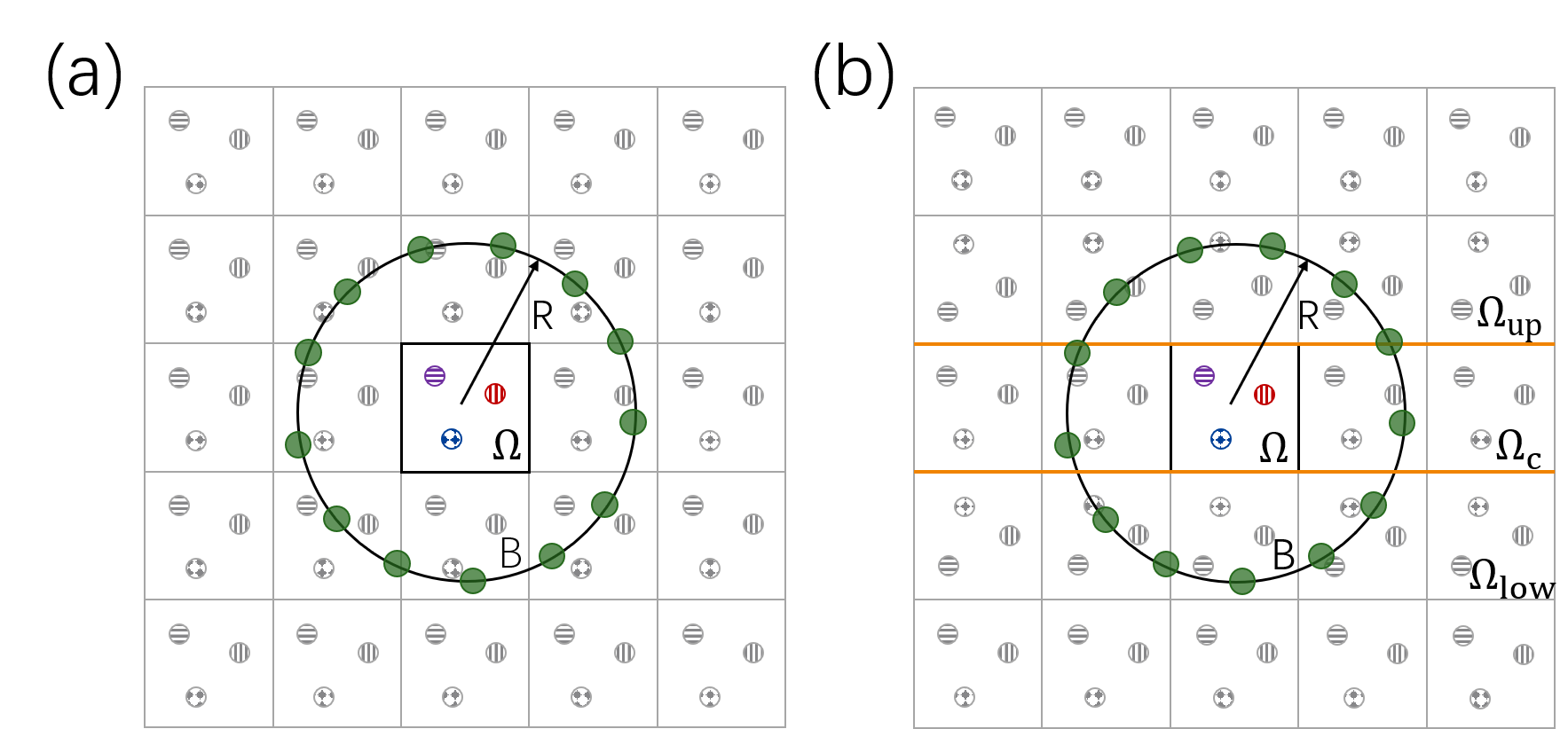}\\
	\caption{Schematic illustration of the HSMA3D (a) and HSMA2D
          (b) for an infinite ionic system under fully/partially
          periodic system. $\Omega$ is the central box, and particles
          within it are the sources. Polarization is approximated by
          the sum of contributions from image charges inside the
          spherical boundary $\partial B$ and the finite point charges
          (green sphere) on $\partial B$. The orange lines indicate
          the locations of dielectric interface.}
	\label{fig:SketchMap}
\end{figure}
A set of $N$ point charges located at positions $\{\bm{r}_{i}=(x_i,y_i,z_i),~i=1,\cdots,N\}$ with
strengths $\{q_i,~i=1,\cdots,N\}$ is placed in the central cuboidal box
$\Omega=[-L_x/2,L_x/2]\times[-L_y/2,L_y/2]\times[-L_z/2,L_z/2]$ which is characterized
by a dielectric constant~$\varepsilon_{\text{c}}$.
For 3D-periodic case, the central simulation box 
is copied along $x$-, $y$-, and $z$-dimensions,
resulting in an infinite number of periodic copies [Fig.~\ref{fig:SketchMap}(a)]. 
In a 2D-periodic system, the simulation box is 
replicated along $x$- and $y$-dimensions where the particles are confined 
by two dielectric boundaries at $z=\pm L_{z}/2$. These interfaces separate 
the entire space $\mathcal{R}$ into three layers which are labeled 
$\Omega_{\text{up}}$, $\Omega_{\text{c}}$, and $\Omega_{\text{low}}$ 
with dielectric constant $\varepsilon_{\text{up}}$, $\varepsilon_{\text{c}}$, 
and $\varepsilon_{\text{low}}$, respectively [Fig.~\ref{fig:SketchMap}(a)].

Let $\Phi(r)$ be the electrostatic potential that satisfies the Poisson's equation within $\Omega$,
\begin{equation}
	-\nabla\cdot\varepsilon(\bm{r})\nabla\Phi(\bm{r})=4\pi\sum_{j=1}^{N}q_j\delta(\bm{r}-\bm{r}_{j}),
\end{equation}  
where $\delta(\bm{r})$ is the Dirac function and the solution can be
expressed as an infinite sum,
\begin{equation}\label{infinitesum}
	\varepsilon_{\text{c}}\Phi(\bm{r})=\sum_{j=1}^{N}\dfrac{q_{j}}
	{|\bm{r}_j-\bm{r}|}+\sum_{j=N+1}^{\infty}\dfrac{q_{j}}{|\bm{r}_j-\bm{r}|},
\end{equation}
where the first term is the direct potential from the source charges and the
second term describes the contribution of infinite images which are
introduced to satisfy the boundary conditions. If the system is fully
periodic, we have 
\begin{equation}
	\varepsilon_{\text{c}}\Phi(\bm{r})=\sum_{\bm{k}}\sum_{j=1}^{N}
	\dfrac{q_{j}}{|\bm{r}_{j}+\bm{k}\circ\bm{L}-\bm{r}|},
\end{equation}
where $\bm{k}=(k_x, k_y, k_z)$ runs over all three-dimensional 
integer vectors and $\bm{L}=(L_x,L_y,L_z)$. Here $\circ$ indicates the element-wise multiplication.  
For a 2D-periodic system with dielectric boundaries, the solution $\Phi_{\text{c}}(\bm{r})$ 
within $\Omega_{\text{c}}$ can be conveniently expressed 
using the image charge construction, leading to,
\begin{equation}
	\begin{split}
\varepsilon_{\text{c}}\Phi_{\text{c}}(\bm{r})=&\sum_{\bm{k}}\sum_{i=1}^{N}\dfrac{q_{i}}{|\bm{r}_{i}+\bm{k}\circ\bm{L}-\bm{r}|}\\
&+\sum_{\bm{k}}\sum_{i=1}^{N}\sum_{\ell=1}^{\infty}\left(\dfrac{q_{i}\gamma_{\ell}^{+}}{|\bm{r}_{i+}^{(\ell)}+\bm{k}\circ\bm{L}-\bm{r}|}+\dfrac{q_{i}\gamma_{\ell}^{-}}{|\bm{r}_{i-}^{(\ell)}+\bm{k}\circ\bm{L}-\bm{r}|}\right),
    \end{split}
\end{equation}
where $\bm{k}=(k_x, k_y, 0)$ represents all two-dimensional 
integer vectors, $\gamma_{\text{up}}=(\varepsilon_{\text{c}}-\varepsilon_{\text{up}})/(\varepsilon_{\text{c}}+\varepsilon_{\text{up}})$ and $\gamma_{\text{low}}=(\varepsilon_{\text{c}}-\varepsilon_{\text{low}})/(\varepsilon_{\text{c}}+\varepsilon_{\text{low}})$ are dielectric mismatches,
$\gamma_{\ell}^{+}=\gamma_{\text{low}}^{\lceil\ell/2\rceil}\gamma_{\text{up}}^{\lfloor\ell/2\rfloor}$, $\gamma_{\ell}^{-}=\gamma_{\text{low}}^{\lfloor\ell/2\rfloor}\gamma_{\text{up}}^{\lceil\ell/2\rceil}$, and the position $\bm{r}_{i\pm}^{(\ell)}$ of the $\ell$th-order image of charge $q_i$ is $\bm{r}_{i\pm}^{(\ell)}=(x_i,y_i,(-1)^{\ell}z_i\pm\ell L_z)$. Here +(-) corresponds to the image charges in $\Omega_{\text{up}}$($\Omega_{\text{low}}$) and the notation $\lceil x\rceil$($\lfloor x\rfloor$) represents the ceil (floor) function.

\subsection{Review of HSMA algorithms}

\begin{figure*}[htbp]
	\centering
	\includegraphics[width=\linewidth]{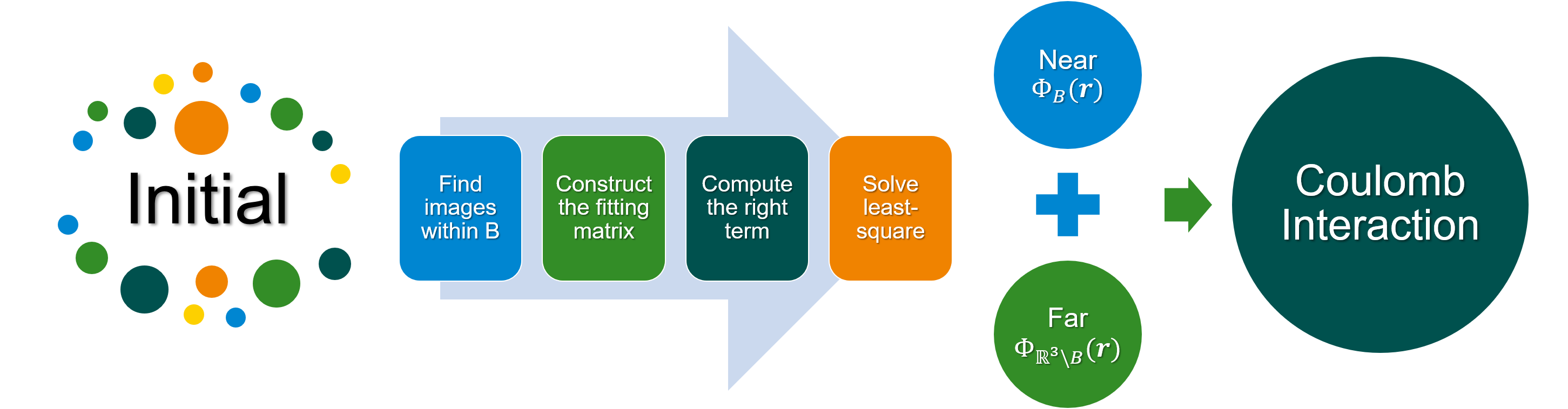}\\
	\caption{The flow chart of the HSMA algorithm.}
	\label{fig:FlowChart}
\end{figure*}

The HSMA algorithms for 3D and 2D periodic systems 
were described in detail in Refs.~[\citenum{zhao2018}] and [\citenum{liang2020}], respectively.
Here we give a brief overview of the most important features of the algorithm.  
A flow chart of the HSMA algorithms is presented in Fig.~\ref{fig:FlowChart}.

First, we find all the images of the source charges within the auxiliary sphere~$B$, which 
includes both periodic images and image charges that are introduced to satisfy dielectric boundary conditions.
We define $\{q_s,~s=1,\cdots,N_s\}$ to be the set of sources and corresponding
image charges within the sphere. Since the electrostatic potential $\Phi_{\mathbb{R}^3\backslash B}(\bm{r})$
of charges outside the sphere~$B$ is a harmonic function in
the domain $\Omega$, the potential~$\Phi(\bm{r})$ can be
split into two parts,
\begin{equation}\label{split}
	\Phi(\bm{r})=\Phi_{B}(\bm{r})+\Phi_{\mathbb{R}^3\backslash
          B}(\bm{r})=\sum_{j=1}^{N_s}\dfrac{q_{j}}{|\bm{r}_{j}-\bm{r}|}+\sum_{n=0}^{P}\sum_{m=-n}^{n}A_{n}^{m}r^{n}Y_{n}^{m}(\vartheta,\varphi)\;,
\end{equation}
where $\Phi_{\mathbb{R}^3\backslash B}(\bm{r})$ is approximated by a truncated
spherical harmonic series~\cite{gumerov2014} with the total number
of spherical basis $N_{b}=(P+1)^2$,
$Y_{n}^{m}$ is the spherical harmonic function of degree $n$ and
order $m$, $\{A_{n}^{m},~n=0,\cdots,P,~m=-n,\cdots,n\}$ is a set of
unknown coefficients. 

Second, we determine the coefficients~$A_{n}^{m}$ of the truncated
spherical harmonic series [Eq.~\eqref{split}] by constructing the fitting matrix and computing the right vector.
In the case of a 3D periodic system, 
we generate $N_m$ ($>N_{b}$) monitoring points which are
nearly uniformly distributed on the circumsphere~$B$
of the central cell $\Omega$.
The coefficients ${A_{n}^{m}}$ are obtained by
solving the least square problem at these points \cite{zhao2018},
\begin{equation}
{A_{n}^{m}}=\text{argmin}\sum_{i=1}^{N_{m}}|\mathcal{L}(\Phi(\bm{r}_{i}))|^2=\text{argmin}\|MA-Y\|_{L^2},
\end{equation} 
where $\mathcal{L}$ is a linear operator on $A_{n}^{m}$, $A$
represents the vector consisting of all the elements of $A_{n}^{m}$, $Y$ is the right vector which depends on the $\Phi_{B}(\bm{r})$ of monitoring points, and $M$ is the mapping matrix which depends on the
locations of monitoring points and the degrees and orders of spherical
basis and can be constructed in the preparation step. 
For a 2D periodic system with dielectric boundary conditions, the above
minimization step cannot be efficiently applied since the dielectric
conditions require information about the exterior domain of the
simulation box $\Omega$. To overcome this issue, we introduce the Dirichlet-to-Neumann (DtN)
boundary condition in Ref.~[\citenum{liang2020}], which transforms the
dielectric conditions into a surface integral that only requires
knowledge of $\Phi_{\text{c}}$ and its derivative in $z$-direction. 
The coefficients $A_{n}^{m}$ are then obtained by
minimizing the residual norm of the DtN boundary condition which is
approximated via the $2$D-dilation formula \cite{occorsio2018}.

Third, we map the spherical harmonic expansion
onto a surface integral over the sphere $B$, which is the core idea of HSMA 
and permits the use of the FMM. 
The main advantages is that such a step avoids the repeated
evaluation of spherical harmonic basis through the recursive relations
due to its less efficient especially for parallel computation.
The surface charge integral is not singular and can be approximated by common numerical quadratures 
such as Fibonacci integration which has a convergence rate of $N_o^{-6}$ with
$N_o$ the number of grid points.

Finally, the FMM can be easily employed to sum over all charges 
within and on the sphere, achieving a truly linear $O(N)$ complexity.
However, direct summation with $O(N^2)$ complexity is also implemented in our package. 

\section{LAMMPS implementation of HSMA}
\label{sec:lammps}

\subsection{Implementation details}

The software is open-source and distributed under GNU General Public
License (GPL). We implement
the HSMA methods described above into LAMMPS as an optional package called
USER-HSMA, providing two new kspace solvers with $O(N)$ complexity
for charged systems with 3D and 2D periodicity. 
The code is available for download from the author's Github page
(\url{https://github.com/LiangJiuyang}),
including documentation and examples.  The software can be installed
via \texttt{make yes-user-hsma} or by directly
copying the files into the LAMMPS main source directory and compiling
the code. 
The required dependent package is the Intel Parallel Studio
\cite{blair2012} (including functions of MPI, OpenMP, MKL
library, and AVX512 instructions). 

The two new kspace styles are called
\texttt{HSMA3D} and \texttt{HSMA2D}, which treats the long-range electrostatic interactions of 3D periodic
system and 2D periodic system with dielectric interfaces,
respectively. The function of \texttt{HSMA3D} is the same as the classical Ewald and the PPPM
in LAMMPS, whereas \texttt{HSMA3D} has the most optimal $O(N)$ complexity by the use of the FMM. 
Furthermore, no alternative approaches for 2D periodic 
system with dielectric interfaces have been implemented into LAMMPS,
and we believe the implementation of \texttt{HSMA2D} will greatly benefit the 
general computational physics community. 
Both solvers allow large scale simulations of molecular
systems and are very general as they can be easily combined with 
other bonded and non-bonded interactions. 
 
\begin{figure*}[htbp]
	\centering
        \includegraphics[width=\linewidth]{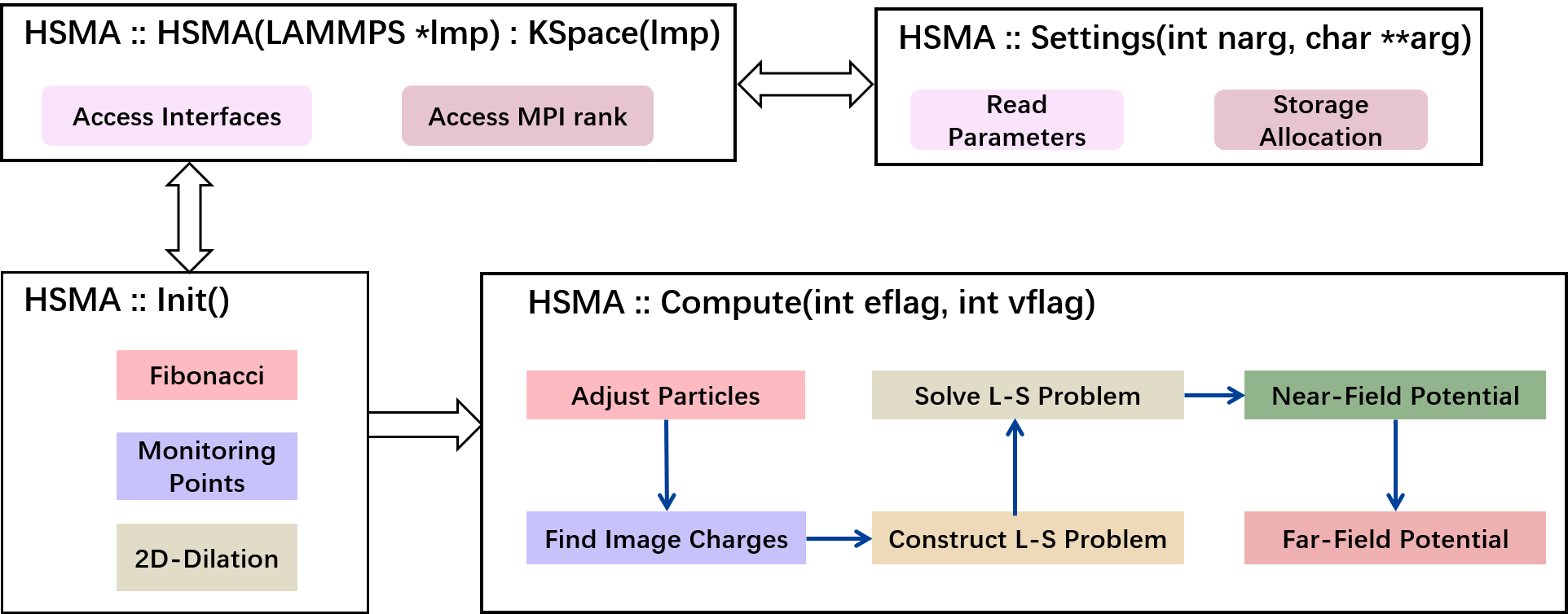}\\
	\caption{The implemented CPU workflow of our HSMA codes in
          LAMMPS.}
	\label{fig:Code_Analysis}
\end{figure*}

The current workflow of
our codes is shown in Fig.~\ref{fig:Code_Analysis}. The interfaces of the information of systems and the MPI
ranks are accessed within the constructor. The ``Settings'' function
includes the computation before the first timestep of a run, including
reading parameters and allocating storages. The ``Init'' function
initializes the calculation before a run, including the generation of
the Fibonacci quadratures, the monitoring points, and the 2D-dilation
quadratures. The ``Compute'' function, which executes at every time
step, involves a complete algorithm flow.

Our current implementation supports massively parallel MD
simulation of large-scale particle systems. We use Intel’s 512-bit
SIMD (AVX-512 architecture) for vectorization implementation outline
in which eight neighbors for double precision (or sixteen for float)
are operated at the same time, and hybrid MPI plus OpenMP are employed for
parallelization. Although the compiler can automatically vectorize
loops, it is sometimes necessary to make changes to the codes which can improve
vectorization efficiency. The pure MPI parallelization is permitted, but
we suggest one use few MPI nodes but more OpenMP threads of each
MPI, because of the saving of communication cost. Our implementation
is consistent with the existing parallelization options provided in
LAMMPS.

The interfaces of
positions, forces, strengths, and other required MD parameters are
obtained from the current class. After execution, the electrostatic
energy, the force of atoms, and the potential are appended to the thermodynamic output
which is useful because one can directly
obtain the required results for different systems without any further
changing in the input file or the source codes, except the
$\emph{kspace\_style}$ command.

\subsection{Setup in LAMMPS}\label{UserInstruction}

We describe the practical usage of \texttt{HSMA3D} and \texttt{HSMA2D} in LAMMPS.
The methods can be readily invoked by setting the following command lines
in LAMMPS,

\texttt{
\\ \detokenize{kspace_style} HSMA3D $\varepsilon_{\text{tol}}$
    $\lambda$ $P$ $N_m$ $F_{w}$ $F$   $f_1$  $f_2$\\
 \detokenize{kspace_style} HSMA2D $\varepsilon_{\text{tol}}$
    $\lambda$ $\gamma$ $P$ $N_w$ $W$ $F_{w}$ $F$   $f_1$  $f_2$ \\ }\\
where the former works for fully periodic systems and the
latter works for partially periodic systems.

The \texttt{HSMA3D} style invokes the HSMA algorithm for 3D periodic system~\cite{zhao2018}.
The specified parameter~$\varepsilon_{\text{tol}}$ is a
double-precision variable that ranges from $0$ (exclude $0$) to
$1$. This parameter will take effect on the accuracy of the FMM. If one
chooses direct summation instead of the FMM, the float64 vector is used
for vectorization when $\varepsilon_{\text{tol}}<10^{-6}$; otherwise,
when $\varepsilon_{\text{tol}}\geq10^{-6}$, the float32 vector is
performed. The parameter~$\lambda$ is a double-precision variable (larger
than $1$) that represents
the relative ratio of the radius of the auxiliary surface and the diagonal length of the simulation box.
A larger $\lambda$ means that more images are included within
the auxiliary surface, thus increases the cost but improves the
accuracy. The parameter~$P$ is a positive integer which indicates that the total number
of spherical basis is $(P+1)^2$. The parameter ~$N_m$ is
the number of monitoring points~$N_m$ which
should be larger than $(P+1)^2$. A recommended choice
is that $N_m=2(P+1)^2$. The parameters $F_w$ and $F$ are two adjacent integers in
the Fibonacci sequence used in spherical integration. In most cases, $F_w=55$ and $F=89$ are accurate enough. 
The parameters $f_1$ and $f_2$ are
integers that are either $1$ or $0$, indicating whether using
the FMM or not when calculating the potential at the monitoring
points and the near-field potential at the source points.

The \texttt{HSMA2D} style invokes the HSMA algorithm for 2D periodic system with dielectric interfaces~\cite{liang2020}.
The definitions of $\lambda$, $P$, $F_w$,
$F$, $f_1$, and $f_2$ are the same as in the \texttt{HSMA3D} style.
Three additional parameters are specified:
the parameter~$\gamma\in[-1,1]$ is the dielectric mismatch, and 
$N_w$ and $W$ are two integers used in the 2D-dilation formula with the total
number of quadratures along $X$ and $Y$ dimensions as
$N_w\lfloor\sqrt{W}\rfloor$. 

The algorithm parameters used to attain different accuracies can
be found in Refs.~\cite{zhao2018,liang2020,liang2021}. For an electrostatic accuracy of $10^{-5}$
which is often adopted for practical simulations, we can safely set

\texttt{
\\ \detokenize{kspace_style} HSMA3D $\varepsilon_{\text{tol}}$
    $1.3$ $10$ $200$ $55$ $89$ $1$ $0$\\
 \detokenize{kspace_style} HSMA2D $\varepsilon_{\text{tol}}$
    $1.6$ $0.94$ $6$ $40$ $4$ $55$ $89$ $0$ $1$\\ }\\


The package requires the use of a matching pair style to
perform short-range pairwise calculations, for example ``lj/cut/omp''
is recommended for a primitive model electrolyte where particles
interact via both long-range Coulomb interactions
and short-range Lennard-Jones (LJ) interactions. 
That said, one should not not use any style which contains ``coul/long''
since the solvers \texttt{HSMA3D} and \texttt{HSMA2D} already contain 
fully electrostatic interactions.


 \section{Accuracy benchmark}
\label{sec:accuracy}

\subsection{Primitive monovalent, divalent, and trivalent electrolytes}
\label{sec:electrolytes-3D}
To confirm the correct functioning of HSMA3D in our package, we
study primitive electrolytes in a 3D periodic system. All ions are modeled as soft spheres of diameter
$\sigma$ that interact through
purely repulsive shifted-truncated LJ potential with
energy coupling $\varepsilon_{\text{LJ}} = k_{\text{B}}T$, where
$k_{\text{B}}$ is the Boltzmann constant and $T$ is the absolute
temperature. The system has dimensions $L_x=25\sigma$, $L_y=25\sigma$, and
$L_z=25\sigma$. We consider three types of model electrolytes, namely
monovalent, divalent, and trivalent electrolytes
containing $1500$, $1000$, $750$ cations and $1500$, $2000$, $2250$
anions, respectively, resulting in a volume fraction $10.05\%$. The ions
are immersed in continuum solvent characterized by a Bjerrum length
$\ell_{\text{B}}=e^2/(4\pi\varepsilon_{\text{c}}k_{\text{B}}T)=3.5\sigma$. We
utilize a time step of $0.01\tau$, where
$\tau=\sigma\sqrt{m_0/k_{\text{B}}T}$ is the unit of time with $m_0=1$
(LJ unit) the ion mass. The simulations are performed in the canonical
ensemble, where the temperature is controlled by a Nos$\acute{\text{e}}$-Hoover
thermostat with damping time $0.01\tau$. Energy minimization proceeds
for $2\times10^5$ time steps, followed by a production period which
occupies $10^7$ time steps. For comparison, we also perform MD simulations
using the PPPM algorithm for which the real-space cutoff is set to
$10\sigma$ and the relative force accuracy is $10^{-5}$.

\renewcommand\arraystretch{1.6}
\begin{table*}[!htbp]\label{Table}
	\centering
	\setlength{\tabcolsep}{2.8mm}{
		\begin{tabular}{|l|cc|cc|cc|cc|}	
			\hline \multirow{2}{*}{} &
			\multicolumn{2}{c|}{Fig.\ref{fig:1_3}(d)} &
			\multicolumn{2}{c|}{Fig.\ref{fig:1_4}(d)} &
			\multicolumn{2}{c|}{Fig.\ref{fig:1_5}(d)} & \multicolumn{2}{c|}{Fig.\ref{fig:1_6}(f)}
			\\\cline{2-9} & Mean & Var & Mean & Var & Mean & Var& Mean &Var
			\\\hline HSMA3D & 0.3561 & 1.2132e-3 & -1.4840 &
			2.3875e-3 & -3.2741 & 3.8337e-3 & -1243057.9 & 431035.2 \\\hline PPPM &
			0.3562 & 1.2384e-3 & -1.4841 & 2.3621e-3 & -3.2738 &
			3.8262e-3& -12431058.1 & 431854.1\\ \hline
	\end{tabular}}
	\caption{The mean value and the variance of total energy produced by HSMA3D and the PPPM for the systems
		in Fig.~\ref{fig:1_3}, Fig.~\ref{fig:1_4}, Fig.~\ref{fig:1_5}, and Fig.~\ref{fig:1_6}.}
\end{table*}

\begin{figure}[!htbp]
	\centering \includegraphics[width=1.0\linewidth]{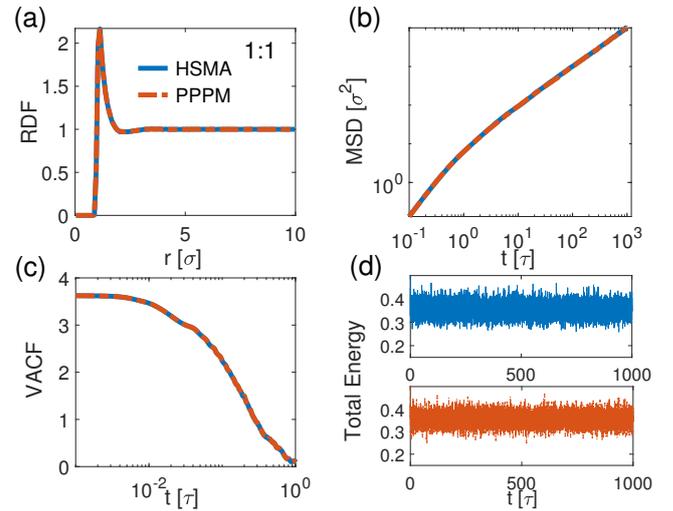}\\
	\caption{Accuracy comparison in 1:1 electrolytes. The
          simulation results from the PPPM (red dash-dot line) and HSMA
          (blue solid line) in electrolytes. (a): The radial
          distribution function of cation-anion in electrolytes. (b):
          The mean square displacement of cation in electrolytes. (c):
          The velocity auto correlation function of cation in
          electrolytes. (d): The total energy (unit: $\varepsilon_{\text{LJ}}$) of the system.}
	\label{fig:1_3}
\end{figure}

\begin{figure}[htbp]
	\centering \includegraphics[width=1.0\linewidth]{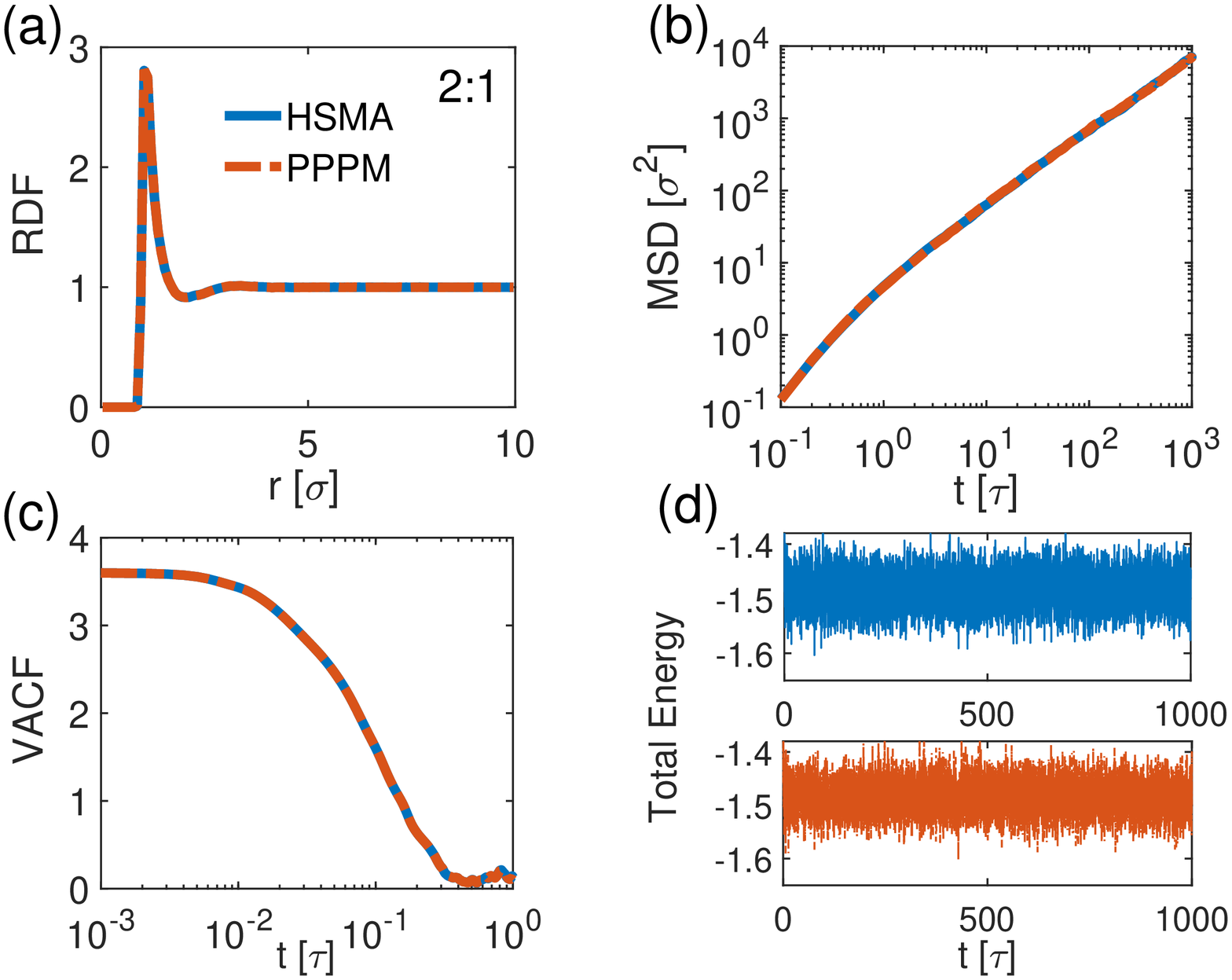}\\
	\caption{Accuracy comparison in 2:1 electrolytes. The
          simulation results from the PPPM (red dash-dot line) and HSMA
          (blue solid line) in electrolytes. (a): The radial
          distribution function of cation-anion in electrolytes. (b):
          The mean square displacement of cation in electrolytes. (c):
          The velocity auto correlation function of cation in
          electrolytes. (d): The total energy (unit: $\varepsilon_{\text{LJ}}$) of the system.}
	\label{fig:1_4}
\end{figure}

\begin{figure}[htbp]
	\centering \includegraphics[width=1.0\linewidth]{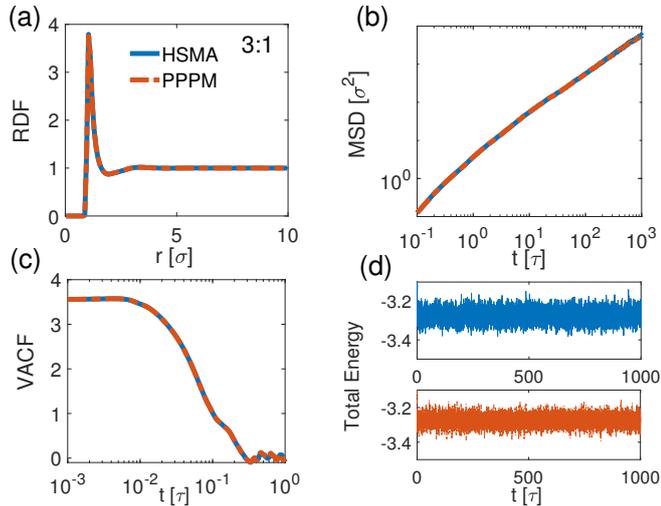}\\
	\caption{Accuracy comparison in 3:1 electrolytes. The
          simulation results from the PPPM (red dash-dot line) and HSMA
          (blue solid line) in electrolytes. (a): The radial
          distribution function of cation-anion in electrolytes. (b):
          The mean square displacement of cation in electrolytes. (c):
          The velocity auto correlation function of cation in
          electrolytes. (d): The total energy (unit: $\varepsilon_{\text{LJ}}$) of the system.}
	\label{fig:1_5}
\end{figure}

We examine both the structural and dynamical properties of 
the model electrolytes, quantified by the radial distribution function
(RDF), the mean square displacement (MSD), the velocity auto
correlation function (VACF), and the total energy evolution
in monovalent (Fig.~\ref{fig:1_3}), divalent (Fig.~\ref{fig:1_4}), 
and trivalent electrolytes (Fig.~\ref{fig:1_5}). 
The HSMA3D and the conventional PPPM in LAMMPS
produce statistically identical RDF $g_\text{ca}$ 
of cations and anions [Fig.~\ref{fig:1_3}(a), Fig.~\ref{fig:1_4}(a), and Fig.~\ref{fig:1_5}(a)].
Notably, the presence of a local minimum in $g_\text{ca}$ 
after the peak shows the charge reversal
in divalent and trivalent electrolyte due to strong electrostatic correlations.
The mean-squared displacement of cations in electrolytes shows the linear dependence [Fig.~\ref{fig:1_3}(b), Fig.~\ref{fig:1_4}(b) and
Fig.~\ref{fig:1_5}(b)] the the velocity auto-correlation
function decays to zero in the long-time limit [Fig.~\ref{fig:1_3}(c), Fig.~\ref{fig:1_4}(c) and
Fig.~\ref{fig:1_5}(c)].
The perfect agreement between HSMA3D and the PPPM
confirms that dynamical properties are properly reproduced, indicating the correct integration of HSMA3D
with other MD integration parts in LAMMPS.
The variation of the total energy of the
whole system during the simulation are presented in Fig.~\ref{fig:1_3}(d), Fig.~\ref{fig:1_4}(d), and
Fig.~\ref{fig:1_5}(d), with the mean value and the variance of
total energy given in Table \ref{Table}, confirming the correct implementation of HSMA3D in LAMMPS.

\subsection{SPC/E bulk water}
We also examine the structure of pure water by employing the classical SPC/E model~\cite{berendsen1987}. 
The cubic simulation box has a dimension of 5.60 nm with periodic
boundary conditions and 17496 atoms were placed within the central cell. 
The equilibrium temperature is set to $T=298K$. 
The equilibration process was carried out for 50 ns in $NPT$ ensemble, 
followed by $NVT$ MD simulation of 100ns for
data collecting with the PPPM and HSMA3D, respectively. Here, six important
quantities, namely the RDF of oxygen-oxygen, oxygen-hydrogen, and
hydrogen-hydrogen, the MSD, the VACF, and the total energy of the
whole system are examined to characterize the structure and dynamics of the SPC/E water. 

\begin{figure}[!htbp]
	\centering \includegraphics[width=\linewidth]{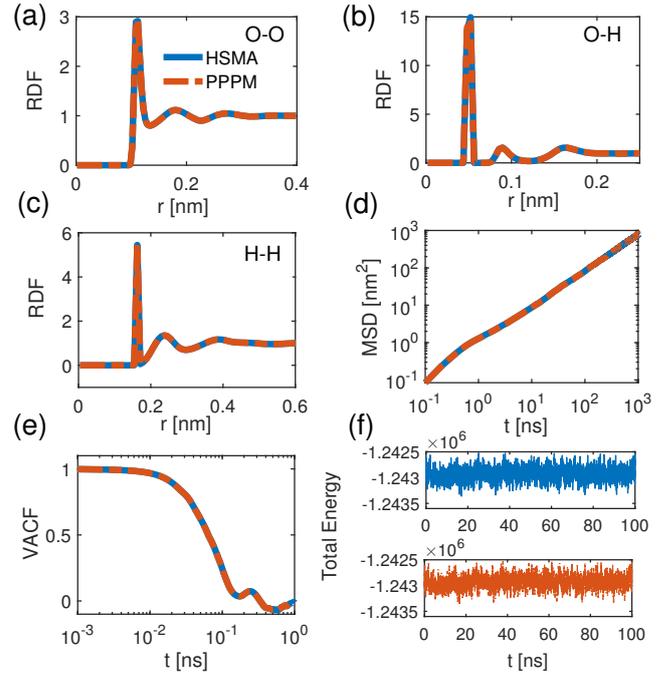}\\
	\caption{Accuracy comparison in SPC/E pure water system. The
          simulation results from the PPPM (red dash-dot line) and HSMA
          (blue solid line) in bulk water. (a-c): The radial
          distribution function of oxygen-oxygen, oxygen-hydrogen, and
          hydrogen-hydrogen in water molecules, respectively. (d): The
          mean square displacement of oxygen in water molecules. (e):
          The velocity auto correlation function of oxygen in water
          molecules. (f): The total energy (unit: Kcal/M)
          of the system. Note that in (f), the mean energy produced by the HSMA and the PPPM is -1243057.9 and
          -12431058.1, and the variance of energy produced by the HSMA and the PPPM is 431035.2
          and 431854.1, respectively. }
	\label{fig:1_6}
\end{figure}

The RDF of oxygen-oxygen,
oxygen-hydrogen, and hydrogen-hydrogen atom pairs are presented in
Fig.~\ref{fig:1_6}(a)-(c), in agreement with previous computational investigations~\cite{mark2001structure}. 
Note, however, that the first peak of the RDF of oxygen-hydrogen and hydrogen-hydrogen in Ref.~\cite{mark2001structure} is missed because the oxygen and hydrogen 
within the same water molecule is not considered.  
The MSD of the oxygen in water molecules is shown in Fig.~\ref{fig:1_6}(d), 
describing the translational motion on different time scales. 
The VACF and the total energy of the whole system, 
are provided in Fig.~\ref{fig:1_6}(e) and Fig.~\ref{fig:1_6}(f).
In all the panels above, we find an excellent agreement between HSMA3D and the PPPM. 

\subsection{Primitive electrolytes between dielectric interfaces}
To demonstrate the accuracy of the HSMA2D for planar dielectric
interfaces in a MD simulation, we study the primitive monovalent and divalent electrolytes
confined between two dielectric interfaces. We assume that the top and
bottom substrates have the same permittivity, i.e.,
$\gamma_{\text{up}}=\gamma_{\text{low}}=\gamma=0.939$. 
The shifted-truncated LJ potential is used for modeling excluded-volume effects
as in Sec.~\ref{sec:electrolytes-3D}.
The system has dimensions $L_x=100\sigma$, $L_y=100\sigma$, and
$L_z=50\sigma$. The monovalent and divalent electrolytes
contain $218$, $109$ cations and $218$, $218$ anions,
respectively. The ions are immersed in continuum solvent characterized
by a Bjerrum length $\ell_{\text{B}}=3.5\sigma$. The ions are confined
by purely repulsive shifted-truncated LJ walls
($\varepsilon_{\text{ion-wall}}=k_{\text{B}}T$;
$\sigma_{\text{ion-wall}}=0.5\sigma$) at $z=-25\sigma$ and
$z=25\sigma$. The MD simulations utilize a time step
$0.005\tau$. Integration proceeds via the time-reversible
measure-preserving Verlet and rRESPA integrators and the simulations
are performed in the $NVT$ ensemble, where the temperature
is controlled by a Nos$\acute{\text{e}}$-Hoover thermostat with damping time
$0.05\tau$. The simulation initially proceeds for $5\times10^5$ time steps for equilibration,
followed by a production period which occupies $10^8$ time steps.

\begin{figure}[!htbp]
	\centering \includegraphics[width=\linewidth]{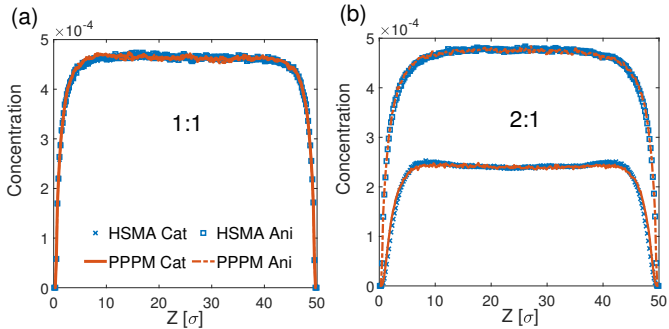}\\
	\caption{Distributions of ion density of (a) a monovalent
          electrolyte and (b) a divalent electrolyte between polarizable interfaces with
          $\gamma=0.939$. We use different symbols and lines for the
          results of different kinds of ions of HSMA2D (solid multiple
          signs for cation and solid squares for anion) and of
          ICM-PPPM (solid line for cation and dash-dotted line for
          anion), and different colors to present different methods
          (blue for HSMA2D, and red for ICM-PPPM). The distributions
          obtained using HSMA2D and ICM-PPPM overlap within statistical
          accuracy. The expected ion depletion near insulating
          interfaces are borne out by the simulations. }
	\label{fig:1_9}
\end{figure}

Ion distributions of monovalent and divalent electrolytes under dielectric confinement 
are presented in Fig.~\ref{fig:1_9}(a) and (b), calculated by HSMA2D along with a comparison to the ICM-PPPM
algorithm~\cite{yuan2021}. As expected, repulsive
polarization results in ion depletion. Such dielectric effects on the
ion distribution are particularly strong for divalent salt.
We observe that the HSMA2D and ICM-PPPM produce statistically identical distributions, confirming
the correct implementation of HSMA2D in LAMMPS.

\section{Performance analysis}
\label{sec:CPU}

The performance comparison of performance between HSMA and the PPPM was
carried out by using the MD engine of LAMMPS (version 7Aug2019) on
primitive electrolytes, all-atom simulation of SPC/E water,
and electrolytes confined by two parallel dielectric interfaces. To access a
fair comparison, the estimated relative force errors $\Delta$ is
chosen as $10^{-5}$. The parameters of the PPPM are chosen
automatically in LAMMPS based on the error estimates~\cite{deserno98b}. 
The parameters of HSMA are chosen
the same as in Sec.~\ref{UserInstruction}. Because HSMA is a package accelerated by a
hybrid MPI plus OpenMP parallelization, the version of the PPPM is chosen as
the optimized multi-threaded version in the USER-OMP package of LAMMPS
for the same acceleration support. Our HSMA package provides two
optimized options. The first is vectorization for directly compute by
using AVX512 instructions. The second is the FMM for achieving a linear
complexity. The former is more efficient for a small system, while the latter is
a general choice for a large system. The publicly available software
package FMM3D \cite{cheng1999,greengard2002} is adopted for the
FMM acceleration.

The computations in this article were performed on the $\pi$ 2.0 cluster
supported by the Center for High Performance Computing at Shanghai
Jiao Tong University. Each CPU node contains two Intel Xeon Scalable
Cascade Lake 6248 (2.5GHz, 20 cores) and 12 $\times$ Samsung 16GB DDR4
ECC REG 2666 memory. The
``intel-parallel-studio/cluster.2020.1-intel-19.1.1'' is used as the
compiler and the LAMMPS is compiled using ``make
intel$\_$cpu$\_$intelmpi''.

We characterize the CPU performance and parallel efficiency of the
implemented package in LAMMPS for a practical simulation study on
SPC/E pure water systems, which is commonly used system in the
all-atom simulation, and monovalent electrolytes confined by two
parallel dielectric interfaces. To reduce the communication cost as
much as possible, we employ few MPI ranks but more OpenMP threads as
recommended in section \ref{UserInstruction}. 

\subsection{Time performance of HSMA3D}

\begin{figure}[!htbp]
	\centering \includegraphics[width=\linewidth]{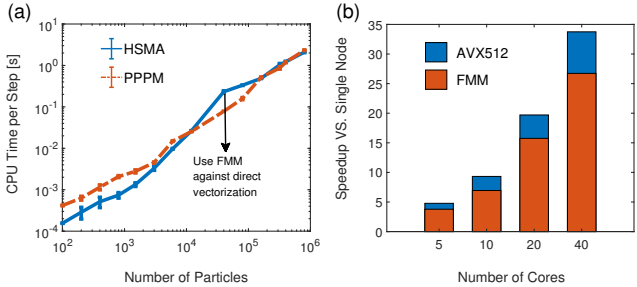}\\
	\caption{Performance analysis in SPC/E pure water system. (a)
          Average CPU time per simulation step as a function of the particle number of pure water systems. $1$ MPI rank with 40
          OpenMP threads are employed for this test. Blue solid line :
          HSMA with $\Delta=10^{-5}$. Red dash-dotted line : PPPM with
          $\Delta=10^{-5}$. Note that for the choice of acceleration
          methods of HSMA, we choose the faster one and label the
          breakeven point in the figure. (b) Speedup on multiple
          compute nodes versus single node calculations for the system
          contains $10^5$ atoms for different ways of acceleration of
          HSMA. Blue : Direct computation with acceleration of
          AVX512. Red : Acceleration by using the FMM. }
	\label{fig:1_7}
\end{figure}

The CPU time of the SPC/E water system as a function
of the number of atoms $N$ for different methods is shown in Fig.~\ref{fig:1_7}.
For this, we use $1$ MPI rank with
$40$ OpenMP threads. The simulations of
the system were conducted for $2000$ steps to estimate the average CPU
time per step. The density of water molecules is fixed to $1~\text{g/c}\text{m}^3$. The real-space cutoff
for LJ potential is set to $10\mathring{\text{A}}$ for both HSMA and
PPPM. Moreover, for the PPPM, we perform a set of simulations
with varying cutoff for real-space Coulomb interaction to obtain the
optimal CPU time. As illustrated in Fig.~\ref{fig:1_7}, HSMA has attractive
performance for various sizes of systems, from $10^2$ to $10^{6}$. When
$N<10^4$, HSMA is faster due to the small prefactor of direct
compute. However, because of the $O(N^2)$ complexity, the performance
of PPPM exceeds HSMA when $N>10^4$. Note that for the choice of
acceleration methods of HSMA, we choose the faster one and label the
breakeven point in the figure. After this breakeven point (about
$3\times 10^4$), the FMM is employed instead of direct vectorization
thus achieves an $O(N)$ linear complexity. The HSMA and PPPM have the
almost same performance when $N>10^5$. Fig.~\ref{fig:1_7}(b) shows the
strong scaling analysis of the HSMA for different ways of
acceleration. The testing system contains $10^5$ atoms. The parallel
efficiency decreases as the number of CPU cores increases, as
expected. Both vectorization and FMM have attractive parallel
efficiency. The parallel efficiency of direct vectorization is
slightly larger than FMM, because the complex data structure contains
more serial computational parts.

In our previous work \cite{zhao2018,liang2020}, the breakeven point
between the FMM and direct computation is about $3000$. However, we find that
the breakeven point moves up by utilizing the vectorization technique for optimization. 
This increment may alter if the FMM code
could also include vectorization. We note
that the above comparison is merely an indication of the relative
efficiency; a complete comparison also requires taking into account
different particle distributions, and other issues \cite{arnold2013}.

\subsection{Time performance of HSMA2D}

The CPU performance and parallel
efficiency of the implemented HSMA2D in LAMMPS are examined for a practical
simulation study on monovalent electrolytes confined by two parallel
plates. To our knowledge,
there are some methods implemented in LAMMPS for dielectric
interfaces~\cite{yuan2021,nguyen2019}
but, no package with the FMM or vectorization optimization exists. Due to
this reason, we compare HSMA2D with the PPPM with slab correction
(PPPM2D) method which is fully optimized in LAMMPS. Note that the
system confined by two parallel plates is a specical case of
dielectric system with no dielectric mismatch, i.e., $\gamma=0$. The
real-space cutoff for PPPM2D is set to $10\sigma$.

\begin{figure}[!htbp]
	\centering \includegraphics[width=\linewidth]{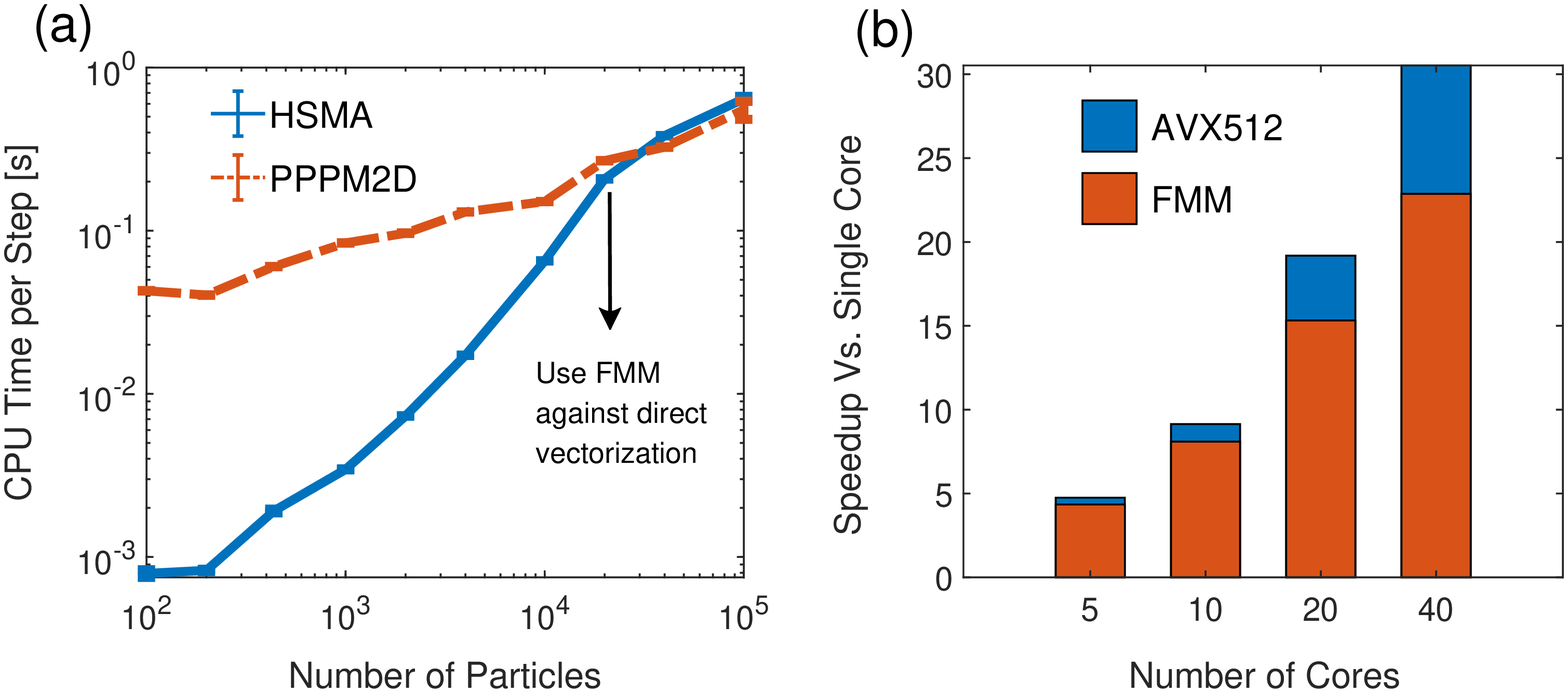}\\
	\caption{Performance analysis in monovalent electrolyte
          confined by two dielectric interfaces. We fix the dimensions
          as $L_x=100\sigma$, $L_y=100\sigma$, and $L_z=50\sigma$ then
          change the ion density of electrolyte. (a) Average CPU time
          per simulation step as a function of the different ranges of
          monovalent electrolyte. $1$ MPI rank with 40 OpenMP threads
          are employed for this test. Blue solid line : HSMA with
          $\Delta=10^{-5}$. Red dash-dotted line : PPPM2D with
          $\Delta=10^{-5}$ and \emph{kspace$\_$modify slab 4}
          command. Note that for the choice of acceleration methods of
          HSMA, we choose the faster one and label the breakeven point
          in the figure. (b) Speedup on multiple compute nodes versus
          single node calculations for system contains $10^5$ atoms
          for different ways of acceleration of HSMA. Blue : Direct
          computation with acceleration of AVX512. Red : Acceleration by
          using the FMM.}
	\label{fig:1_8}
\end{figure}

The average CPU time per simulation
step as a function of the particle number of monovalent
electrolyte is shown in Fig.~\ref{fig:1_8}. Note that we fix the box size and alter the density of
electrolytes in this test. The simulations of the system were
conducted for 2000 steps to estimate the average CPU time per
step. $1$ MPI rank with $40$ OpenMP threads is performed for this
test. Note that for the choice of acceleration methods of HSMA, we
choose the faster one and label the breakeven point (about $2\times
10^4$) in the figure. As shown in Fig.~\ref{fig:1_8}(a), HSMA2D is
significantly faster (about $1.5$ magnitude) at low particle numbers,
but beyond the break-even points of $N\approx 3\times10^{4}$ (occpies
about $3.1\%$ ion volume fraction), the PPPM2D and the HSMA2D have
nearly the same performance. This is because the direct
vectorization is used when $N\leq2\times 10^4$, resulting in high efficiency
but $O(N^2)$ complexity. We do not test larger systems ($N>10^{5}$)
because of the more memory allocation than the accessible memory per
node. It can be predicted that the HSMA2D with the FMM acceleration will
have more advantages if the system is larger than $10^5$ due to the
linear scaling. Fig. \ref{fig:1_8}(b) shows the strong scaling
analysis of the HSMA2D for different ways of acceleration. The testing
system contains $10^5$ atoms. The parallel efficiency decreases as the
number of CPU cores increases, as expected. Both vectorization and the FMM
have attractive parallel efficiency and a slight loss than the results
in fully periodic system as shown in Fig. \ref{fig:1_7}(a) thanks to
the complicated calculation of the 2D-dilation formula.
  
\section{Conclusion}
We have implemented HSMA methods for modeling electrostatic interactions of
fully periodic and partially periodic systems in LAMMPS. We have validated
our implementation by comparing the results of primitive
electrolytes and all-atom pure water to those obtained by other numerical 
methods. The codes are released as an open-source USER-HSMA package which can be easily
compiled into LAMMPS. We have
compared the CPU performance and the parallel efficiency of the
implemented HSMA for practical simulations, showing the attractive
performance and good strong scaling. The HSMA for fully/partially
periodic systems is very efficient for both small and large
particle numbers. This implementation allows efficient and accurate coarse-grained and
all-atom molecular dynamics simulations of 
a broad variety of charged systems, such as, 
polyelectrolytes, ionic liquids, proteins, or membranes.

\section{Acknowledgments and Competing interests}
The authors acknowledge the Center for High Performance Computing at
Shanghai Jiao Tong University for the computing resources. The work of
J. Liang and Z. Xu was partially supported by NSFC (grant
Nos. 12071288) and Shanghai Science and Technology
Commission (grant No. 20JC1414100). All the authors have declared no
competing interest.

\bibliographystyle{elsarticle-num} \bibliography{journals,Jiuyang,chargeregulation,polyelectrolyte,simu,colloids,dielectrics,electrolyte,slab}

\end{document}